\documentclass[aip,preprint,jcp]{revtex4-1}
\usepackage{amsmath,amssymb,amsfonts}
\usepackage{graphicx}
\usepackage{dcolumn}
\usepackage{bm}
\usepackage{color}

\begin{document}

\title{Anomalous law of cooling}

\author{Luciano C. Lapas}
\email{luciano.lapas@unila.edu.br}
\affiliation{Universidade Federal da Integra\c{c}\~{a}o Latino-Americana, Caixa Postal 2067, 85867-970 Foz do Igua\c{c}u, Paran\'a, Brazil}

\author{Rogelma M. S. Ferreira}
\email{rogelma.maria@gmail.com}
\affiliation{Centro de Ci\^{e}ncias Exatas e Tecnol\'{o}gicas, Universidade Federal do Rec\^{o}ncavo da Bahia, 44380-000 Cruz das Almas, Bahia, Brazil}

\author{J. Miguel Rub\'{i}}
\email{mrubi@ub.edu}
\affiliation{Departament de F\'{\i}sica Fonamental, Facultat de F\'{\i}sica, Universitat de Barcelona, Av. Diagonal 647, 08028 Barcelona, Spain}

\author{Fernando A. Oliveira}
\email{fernando.oliveira@pq.cnpq.br}
\affiliation{Instituto de F\'{\i}sica and Centro Internacional de F\'{\i}sica da Mat\'{e}ria Condensada, Universidade de Bras\'{\i}lia, Caixa Postal 04513, 70919-970 Bras\'{\i}lia, Distrito Federal, Brazil}

\begin{abstract}
We analyze the temperature relaxation phenomena of systems in contact with a thermal reservoir that undergo a non-Markovian diffusion process. From a generalized Langevin equation, we show that the temperature is governed by a law of cooling of the Newton's law type  in which the relaxation time depends on the velocity autocorrelation and  is then characterized by the memory function. The analysis of the temperature decay reveals the existence of an anomalous cooling in which the temperature may oscillate. Despite this anomalous behavior, we show that the variation of entropy remains always positive in accordance with the second law of thermodynamics.
\end{abstract}



\maketitle

\section{Introduction}\label{sec:intro}

For more than one century, several efforts have been made to study the physical mechanisms that govern the relaxation dynamics of nonequilibrium systems\cite{Kohlrausch854,Kohlrausch863,Mittag905,Zwanzig70}. A non-exponential behavior~\cite{Vainstein06} of the correlation functions has been observed in systems such as supercooled colloidal systems~\cite{Rubi04}, glasses and granular material~\cite{Holek04,Vainstein03}, hydrated proteins~\cite{Peyrard01}, growth~\cite{Calaiori01,Mello01,Aarao06}, plasmas~\cite{Ferreira91}, and disordered vortex lattices in
superconductors~\cite{Bouchaud91}. These systems present similar relaxation features to those found in anomalous diffusive systems. The first attempt to explain such a temperature relaxation was performed by Newton who proposed the relaxation law
\begin{equation}
\frac{d\Delta T(t)}{dt} = - \beta \Delta T(t) \text{,} \label{New}
\end{equation}
where $\Delta T(t)=T(t)-T_R$, with $T(t)$ the temperature of the system and $T_R$ the reservoir temperature. When the relaxation time $\beta^{-1}$ is a constant, this difference decays exponentially from $\Delta T(0)$ to zero.

Few hundred years after Newton, there are still many open questions concerning the temperature relaxation in complex system in which diffusion is anomalous~\cite{Morgado02,Bouchaud90,Ferreira12,Metzler00,Zapperi01,Filoche08,Srokowski00,Costa03,Lapas07,Lapas08,Siegle10}. The study of systems with long-range memory reveals some physical phenomena that are still not well understood. At first sight, the presence of oscillations in the relaxation function raises some uncertainties about the validity of the second law of thermodynamics such as occurs in the temporal relaxation patterns of complex systems~\cite{Metzler99}. 

In the present work, we analyze the temperature relaxation in systems exhibiting anomalous diffusion whose dynamics can be described by a Generalized Langevin Equation (GLE). We show that the presence of memory effects is responsible for an anomalous cooling in which the temperature of the system may oscillate.

The paper is organized as follows. In Sec.~\ref{sec:temp_rel_model}, we obtain the velocity-velocity correlation function for an ensemble of particles and derive a general expression for the temperature, considering a gas of correlated particles. We show that Newton's law of cooling exhibits an anomalous behavior, becoming more pronounced as the non-Markovian effects are more important. In Sec.~\ref{sec:entropy}, we perform an analysis for entropy and scaling in order to show that, according to the second law of thermodynamics, the entropy production is non-negative even for temperature oscillations and that a wide range of diffusive processes exhibits universal behavior. Finally, in Sec.~\ref{sec:conclusions}, we present our main conclusions.

\section{Temperature relaxation model}\label{sec:temp_rel_model}

The GLE is a stochastic differential equation that has been frequently used to model systems driven by colored random forces~\cite{Toussaint04,Dias05,Rahman62,Yulmetyev03,Vainstein05,Bao03,Bao06}. These systems are commonly found in the dynamics of polymeric chains~\cite{Toussaint04,Dias05}, metallic liquids~\cite{Rahman62}, Lennard-Jones liquids~\cite{Yulmetyev03}, the diffusion of spin waves in disordered systems~\cite{Vainstein05}, and ratchet devices~\cite{Bao03,Bao06}. For the velocity $v(t)$, this equation can be written as,
\begin{equation}
m\frac{dv(t)}{dt} = - m\int_0^t \Gamma(t - t')v(t')dt' + \xi(t)\text{,} \label{eq1}
\end{equation}
where $\Gamma(t)$ is the retarded friction kernel of the system or memory function. For this reason, the GLE can model the non-Markovian movement of a particle of mass $m$ with a time-dependent friction coefficient. By assuming a generalized stochastic process, the internal random force $\xi(t)$ is subject to the conditions $\langle\xi(t)\rangle = 0$ and $\langle\xi(t)v(0)\rangle = 0$, and Kubo's fluctuation-dissipation theorem (FDT) can be formulated as~\cite{Kubo66,Kubo85,Balucani03}
\begin{equation}
C_{\xi}(t-t') = \langle\xi(t)\xi(t')\rangle = m^2\langle v^{2}\rangle_{eq} \Gamma(t)\text{,} \label{eq2}
\end{equation}
where $C_{\xi}(t)$ is the correlation function for $\xi(t)$ and the angular brackets denote  average over the ensemble of particles~\cite{Kubo85,Lapas07}, \textit{i.e.}, an averaging over the probabilities of all possible values of the quantity $\xi(t-t')$ at independent times $t$ and $t'$. According to the ergodic hypothesis, this statistical average is equivalent to a time average, whereas the mechanical system goes to equilibrium. In addition, since the system reaches the thermal equilibrium, one can assume that $\left\langle v^{2}\right\rangle_{\mathrm{eq}} \propto T$ is related to the equilibrium energy with the thermal bath at absolute temperature $T$. Thus, the strength of correlation function of the stochastic noise is related to the absolute temperature of the heat bath (dissipation and fluctuations relate to the same source) and the memory kernel satisfies $\lim_{t\rightarrow \infty}\Gamma(t)=0$; if the dissipation and fluctuations originate from different sources, the FDT does not hold~\cite{Costa03,Madrid03}.

In this framework, one can consider a thermal reservoir of harmonic oscillators, in which the noise in Eq. (\ref{eq1}) can be obtained as in Ref.~\cite{Vainstein06}
\begin{equation}
 \xi(t) = \int \sqrt{2k_BT g(\omega)} \cos [\omega t + \phi(\omega)] d \omega\text{,} \label{noise}
\end{equation}
where $g(\omega)$ is the noise spectral density and $\phi(\omega)$ a random phase, which is defined in the interval $ 0 < \phi(\omega) < 2 \pi$. From the FDT, Eq. (\ref{eq2}), one obtains
\begin{equation}
\Gamma(t)=  \int g( \omega) \cos(\omega t) d\omega\text{.} \label{Gamman2}
\end{equation}

We recognize a variety of diffusive systems which do not obey the classical Einstein diffusion theory (normal diffusion). In contrast to the linear behavior observed from the normal diffusion, the anomalous diffusion is characterized by a nonlinear behavior at long times stemmed from mean square displacement $\left\langle x^{2}\right\rangle$. We can write $\left\langle x^{2}\right\rangle \sim t^{\alpha}$, where $\alpha$ is the diffusion exponent. When $0<\alpha<1$ the system is considered as subdiffusive, normal for $\alpha=1$, and superdiffusive for $1<\alpha \leq 2$; the ballistic diffusion occurs for $\alpha=2$. In these cases, one verifies that there is a general relationship between the memory function and the diffusion exponent in Laplace space~\cite{Morgado02}
\begin{equation}
\lim_{z\rightarrow0}\widetilde{\Gamma}(z)\propto z^{\alpha-1}, \label{eq:limz}
\end{equation}
where $\widetilde{\Gamma}(z)$ is the Laplace transform of the memory function $\Gamma(t)$. The above expression is also equivalent to the relationship of effective friction, $\gamma^\ast=\int_{0}^{\infty}\Gamma(t)dt$, which is null for all superdiffusive motions $1<\alpha\leq2$. Note that for superdiffusion $\alpha > 1$ the effective friction $\gamma^\ast = \lim_{z \rightarrow 0} \widetilde{\Gamma}(z)$ is null.

A self-consistent equation for the velocity autocorrelation function $C_v(t) = \langle v(t) v(0)\rangle$ can be obtained by multiplying Eq.~(\ref{eq1}) by $v(0)$ and taking the ensemble average, yielding 
\begin{equation}
\frac{dR(t)}{dt} = -\int_0^t \Gamma(t - t')R(t')dt' \text{,} \label{eq6}
\end{equation}
where the normalized correlation function is $R(t) \equiv C_v(t)/C_v(0)$. This equation can be solved in order to analyze the diffusive behavior of the system.  The velocity and its mean square value are given respectively by
\begin{equation}
v(t) = v(0)R(t)+\int_0^t R(t - t')\eta(t')dt \label{ve}
\end{equation}
and 
\begin{equation}
\langle v^{2}(t)\rangle = \langle v^{2} \rangle_{eq}  + R^2(t)\left[ \langle v^{2}(0)\rangle - \langle v^{2} \rangle_{eq} \right]\text{,}  \label{ve2}
\end{equation}
where the subscript $eq$ indicates the ensemble average at the reservoir temperature with $\langle v^{2}(0)\rangle$ being the average over the initial conditions. For the sake of simplicity, since all interactions among particles are implicit in $R(t)$, we can consider a gas of noninteracting particles with constant specific heat $C_V= fk_B/2$, where $f$ is the number of degrees of freedom and $k_B$ is the Boltzmann constant. In this sense, we can associate the kinetic temperature (hereinafter called temperature)~\cite{Lapas07} with kinetic energy to obtain from Eq. (\ref{ve2}) the temperature evolution
\begin{equation}
T(t) = T_{R} + R^2(t)\left[T_0 - T_{R}\right]\text{,} \label{Temp}
\end{equation}
where $T_0$ stands for the initial temperature of the system. Temperature then depends on the normalized velocity correlation which in turns rely on the memory function related to the nature of the environment in which particle moves.
 
The temperature variation $\Delta T(t) = T(t) - T_{R}$ is a solution of  Eq. (\ref{New}) with a relaxation time given by 
\begin{equation}
 \beta = -2 \frac{d }{dt}\ln{[R(t)]}\text{.} \label{beta}
\end{equation}
The cases in which the correlation is $R(t) = \exp{(- \gamma t)}$, with $\beta = 2\gamma$,  correspond to Newton's cooling law. In many cases, however, the normalized correlation function is neither an exponential nor a stretched exponential~\cite{Vainstein06}. Thus, the relaxation time $\beta$ is a function of time and Newton's law is no longer valid.

One should note that the long-time behavior of $R(t)$ corresponds to small values of $z$ in the Laplace transform. Indeed, from the Laplace transform applied to Eq.~(\ref{eq6}), one obtains
\begin{equation}
\widetilde{R}(z)=\frac{1}{z + \widetilde{\Gamma}(z)}\text{.} \label{eq7}
\end{equation}
By considering the usual memory function 
\begin{equation}
\Gamma(t)=  \frac{\gamma}{\tau} \exp{ (-t/\tau)}\text{,}\label{Gamman}
\end{equation}
Eq. (\ref{eq1}) becomes, for small $\tau$, the Langevin equation without memory, {\it i.e.}, $\lim_{\tau \rightarrow 0} \Gamma(t) =  2\gamma \delta(t)$. In addition, the Laplace transform of Eq. (\ref{Gamman}) gives $\widetilde{\Gamma}(z)= \gamma/(\tau z+1)$. For this particular memory, Eq. (\ref{eq7}) can be solved analytically as
\begin{equation}
R(t) =  \frac{\tau}{\rho} \left[ Z_{+}\exp{ (Z_{-}t)} - Z_{-}\exp{ (Z_{+}t)} \right] \text{,}\label{Ra}
\end{equation}
where $\rho =\sqrt{1-4\gamma \tau}$ and $Z_{\pm} = (-1 \pm \rho)/(2\tau)$ are poles of $\tilde{R}(z)$. 

\subsection*{A measure for the degree of non-Markovian behavior}
\begin{figure}
\centering \includegraphics[width=0.5\linewidth]{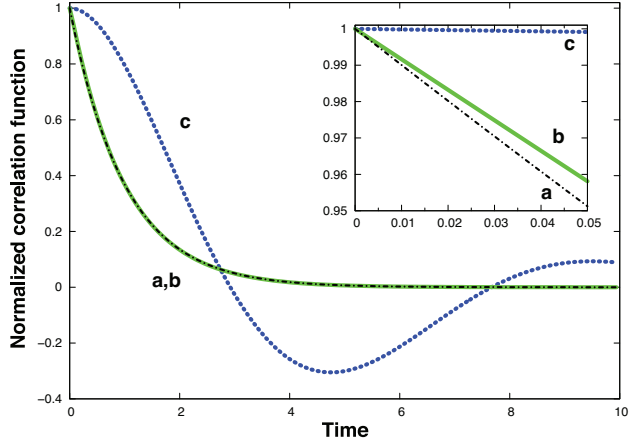}
\caption{\label{fig1}(color online). Numerical results for normalized correlation function $R(t)$ as a function of time $t$ in arbitrary units. For $\gamma=1$, ({\bf a}) the dash-dot line results from an exponential decay, $R(t)=\exp{(-\gamma t)}$. The numerical results are obtained with the use of Eq. (\ref{Ra}) for ({\bf b})  $\tau=0.01$ and ({\bf c}) $\tau=2$. In addition, ({\bf c}) we scale the axes as $\beta$ ($\times 10^{-1} $) and $t$ ($\times 10^{-2}$). The inset highlights the difference between the exponential adjustment and the curve {\bf b} near the origin. In both curves $\beta(0)=0$.}
\end{figure}

The behavior of the correlation depends on the characteristic relaxation time $\tau_0 \equiv (4 \gamma)^{-1}$. For $\tau < \tau_0$, $\rho$ is real and the correlation decays exponentially, as follows from Fig.~\ref{fig1}({\bf a}) and Fig.~\ref{fig1}({\bf b}). For $\tau > \tau_0$, an oscillatory behavior takes place, see Fig.~\ref{fig1}({\bf c}). 
We assume $\beta(0)=0$ in both curves.
The dash-dot line, Fig.~\ref{fig1}({\bf a}), corresponds to $R(t)=\exp{(-\gamma t)}$ with $\gamma=1$. The behavior of the correlation function in Eq. (\ref{Ra}) is illustrated in Fig.~\ref{fig1}. The relaxation time $\tau_0$ thus establishes a threshold for the appearance of non-Markovian effects in the system. As can be observed in Fig.~\ref{fig1}, for small value of $\tau$ one can observe that $R(t)$ is very close to the exponential, except at the origin where the derivative of the response function vanishes. Strictly speaking, as required by Eq. (\ref{eq6}), the response function
\begin{equation}
\frac{dR(t)}{dt} = -\frac{\gamma}{\rho} \left[ \exp{ (Z_{+}t)} - \exp{ (Z_{-}t)} \right]\text{,} \label{dRa}
\end{equation}
is null at the origin. Note that the above equation can be obtained directly from Eq. (\ref{Ra}). In addition, it is possible to obtain $\beta$ and all the dynamics of the system from Eqs. (\ref{Ra}) and (\ref{dRa}).

\begin{figure}
\centering \includegraphics[width=0.5\linewidth]{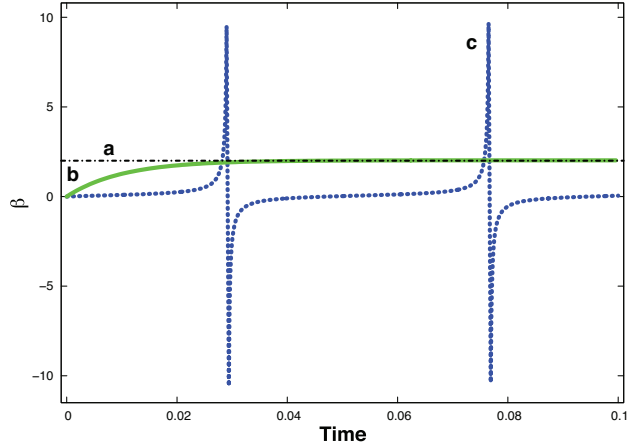}
\caption{\label{fig2}(color online). Numerical results for $\beta$ as a function of time $t$. We use the same data of Fig.~\ref{fig1}. ({\bf a}) $\beta$ is constant only for the exponential decay. ({\bf b}) For small $\tau$, it approaches to the constant value very fast. ({\bf c}) For $\tau=2$, it shows a fast change function with zeros and poles.}
\end{figure}

In Fig.~\ref{fig2}, we plot $\beta(t)$ as a function of time using the same parameters as in Fig.~\ref{fig1}. Data are obtained using Eqs. (\ref{beta}), (\ref{Ra}), and (\ref{dRa}). The dash-dot line, Fig.~\ref{fig2}({\bf a}), corresponds to the case $\beta=const.$, proper of Newton's law. In Fig.~\ref{fig2}({\bf b}), $\beta$ approaches rapidly to the constant $2\gamma$. In Fig.~\ref{fig2}({\bf c}), $\beta$ exhibits oscillations which show that when $\tau$ increases Newton's law is not longer valid. For $\tau <\tau_0$, by using Eq. (\ref{Ra}), one obtains
\begin{equation}
\lim_{t\rightarrow\infty} \beta(t) = \frac{2\gamma}{1-\rho}\text{.} \label{bef}
\end{equation}
For $\tau >\tau_0$, $\beta(t)$ does not reach a fixed value, rather it oscillates as in Fig. ~\ref{fig2}({\bf c}).

\begin{figure}
\centering \includegraphics[width=0.5\linewidth]{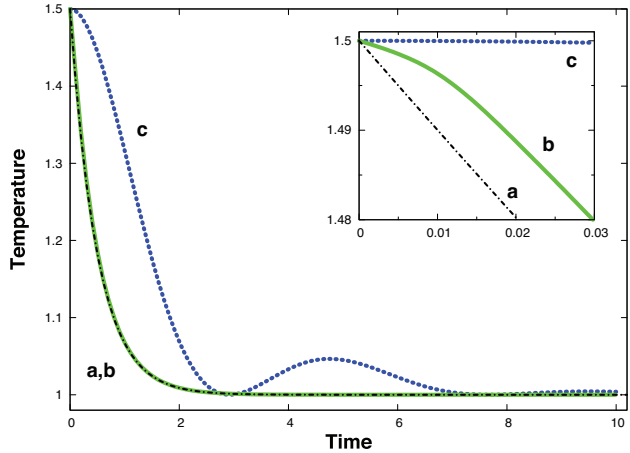}
\caption{(color online). Temperature $T$ as a function of time $t$, which is obtained from Eq. (\ref{Temp}). We assume the reservoir temperature as $T_{R} =1$ and the gas initial temperature $T_0=1.5$. The exponential adjustment ({\bf a}) fits in remarkable well for ({\bf b}) $\tau=0.01$, except near the origin, while for large values ({\bf c}) $\tau=2$ it shows an oscillatory behavior. The inset shows the behavior near the origin.}\label{fig3}
\end{figure}

In Fig.~\ref{fig3}, the temperature (in terms of mean kinetic energy) for an external force-free case is represented as a function of time using the same parameters as previously. We use Eq. (\ref{Temp}) and $R(t)$ as in Fig.~\ref{fig1}, assuming the reservoir temperature $T_{R}=1$ and the gas initial temperature as $T(0)=1.5$. In either cases one has the normal process of cooling, Fig.~\ref{fig3}({\bf a}) and Fig.~\ref{fig3}({\bf b}), while for $\tau>\tau_0$ the temperature shows some oscillations before it reaches equilibrium, as shown in Fig.~\ref{fig3}({\bf c}).
 
\section{Entropy and scaling analysis}\label{sec:entropy}
Let us consider a nonequilibrium gas of Brownian particles characterized by the probability density function (PDF) $P(\mathbf{\Omega};t)$, where $\mathbf{\Omega}=(x,v)$, being $x$ and $v$ the position and the velocity of a Brownian particle. Assuming a homogeneous bath, entropy variation is given by the Gibbs postulate at a local level in the phase space,
\begin{equation}
\delta S = \frac{\delta E}{T} - k_B\int P (\mathbf{\Omega},t)  \ln \delta P(\mathbf{\Omega};t) d\mathbf{\Omega}\text{,}\label{eq:gibbs}
\end{equation}
where $E$ is the energy of the particles~\cite{Madrid94,Rubi99,Vilar01}. The Eq. (\ref{eq:gibbs}) consists of two parts: I. Cooling process (first term on the right side) and II. Diffusive process (second term on the right side). Assuming an instantaneous temperature, $T(t)>T_R$, since the system undergoes an adiabatic process the energy change per unit of time in each part is solely due to the flow of energy, $(1/T(t))\partial E/\partial t=-\Phi_B/T+\Phi_R/T_R$, where $\Phi_B$ and $\Phi_R$ is the heat flow of the Brownian system and thermal reservoir, respectively. According to the first law, the heat gained by one part is equal to the heat lost by another, {\it i.e.}, the rate of heat flow or the heat current $\Phi_B=-\Phi_R=\partial Q/\partial t$, which is given by the heat conduction laws. Regarding the framework of Fourier's law of heat conduction, $\partial Q/\partial t=\kappa (T-T_{R})$, where $\kappa$ is the coefficient of heat conductivity, the thermodynamic flow $\partial Q/\partial t$ is driven by the thermodynamic force $(1/T_{R}-1/T)$~\cite{Prigogine98,deGroot84}. By assuming a quasi-static process, one can consider that the temperature is a function of time to obtain
\begin{equation}
\frac{1}{T(t)}\frac{\partial E}{\partial t}=\kappa \frac{\Delta T^2}{T(t)~T_{R}} \geq 0\text{,}\label{eq:dEdt3}
\end{equation}
independently of the value which $R(t)$ assumes if one considers Eq. (\ref{Temp}). Therefore, the second law holds for the entire system at any time interval. 

To obtain the analytic expression for $P(\mathbf{\Omega};t)$ is in general a difficult task, in many cases unachievable. By the principle of conjunctive probability, the PDF can be written as $P(x,v)=F(x)G(x,v)$. This form holds when the system is far from equilibrium, in whose case the distribution is not a Gaussian. Deviations from a Gaussian behavior, measured by the non-Gaussian factor $\Lambda(t)=\Lambda(0)h(t)$, where $h(t)\propto R^4(t)$, were analyzed in Ref.~\cite{Lapas08}. This relation shows that when the distribution is initially Gaussian, $\Lambda(0)$ is zero, consequently, $\Lambda(t)=0$ and the distribution remains Gaussian for any value of time. Moreover, when the system is described by a non-Gaussian distribution, it evolves rapidly to a Gaussian since $R^4(t)$ goes to zero sufficiently fast. In addition, the equation of motion, Eq. (\ref{eq1}), is not space dependent, as well the temperature $T(t)$ is a function of time only, not involving $x$-space. Since there is no space gradient, the system is homogeneous and consequently the velocity distribution will be space independent. Under this condition, one can consider the factorization $P (\mathbf{\Omega}; t) = F(x)G(v)$, where both functions $F(x)$ and $G(v)$ are Gaussian~\cite{Adelman76,Vilar12}. This factorization still holds when the system is in a local quasi-equilibrium state in which the Gaussian nature of the distribution is preserved for a temperature $T(t)$ instead of $T_R$, a situation compatible with our definition of kinetic temperature via the equipartition law. The distributions $F(x)$ and $G(v)$ have variances $\sigma^2_x(t)=2D(t)t$ and $\sigma^2_v(t)=k_BT(t)/m$, respectively. Local quasi-equilibrium states have been found in the relaxation of glassy systems and granular flows~\cite{Holek04}. Under adiabatic cooling in time (Eq. (\ref{eq:dEdt3}) is null), Eq. (\ref{eq:gibbs}) becomes
\begin{equation}
\Delta S(t) = S(t)-S(t_0)= k_B\ln{\frac{\psi(t)}{\psi(t_0)}} \text{,}\label{SG}
\end{equation}
where
\begin{equation}
\psi(t)=\sigma^2_v(t) \sigma^2_x(t)\text{.}\label{Sigma}
\end{equation}
Since one cannot infer about system response for times smaller than the collision time ({\it i.e.}, delta function is not allowed at origin), both in the Boltzmann and in the Langevin formalism there is a minimum time physically accessible. Consequently we start counting with $t_0 \neq 0$, which in our units yields $\langle v^2(t_0)\rangle=k_B T(t_0)/m$ and  $\langle x^2(t_0)\rangle=2D(t_0)t_0$.

Similar to other models of anomalous diffusion, where unconfined Langevin equations are formally equivalent to a diffusion equation with a time-dependent diffusion coefficient~\cite{Fulinski13,Jeon14,Metzler14}, the diffusion coefficient can be defined as~\cite{Kubo85}
\begin{equation}
D(t) = \int_0^{t} C_v(t')dt'=\frac{k_BT(t)}{m}\phi(t)\text{.}\label{dif}
\end{equation}
with
\begin{equation}
\phi(t) = \int_0^{t} R(t')dt'\text{.}\label{phi}
\end{equation}
In this context, it is important to show that the correlation exhibits peculiar behavior when non-Markovian effects become more relevant. Thus, we rewrite Eq. (\ref{Sigma}) as
 \begin{equation}
 \psi(t) =  \frac{2k_B}{m}T^2(t)\phi(t)\text{,} \label{Rho}
\end{equation}
which is a growing function of $t$. 

Following a recent general result for diffusion~\cite{Ferreira12}, we can use the scaling transformation
\begin{equation}
z \rightarrow \lambda(t)/t \label{zt}
\end{equation} 
in order to obtain $\lambda$, $R(t)$,  and the diffusion coefficient
\begin{equation}
  D(t)= \frac{k_BT(t)}{m}\widetilde{R}\left(z=\lambda(t)/t\right)= \frac{k_BT(t)}{m}\frac{t}{\lambda + t \widetilde{\Gamma}\left(\lambda(t)/t\right))}\text{.}  \label{eq9b}
\end{equation}

Let us consider an extreme case, the ballistic diffusion. In the ballistic regime the effective friction $\gamma^\ast=\lim_{z \rightarrow 0}\widetilde{\Gamma}(z) \rightarrow 0$, {\it i.e.}, the entropy production is a minimum~\cite{Lapas07}. For example, we can consider the spectral density
\begin{equation}
            g(\omega) = \left\{
                       \begin{array}{ll}
                        \frac{A}{\omega_1-\omega_2} \qquad  \omega_2 \leq \omega \leq \omega_1 \\
                        0 \qquad \text{otherwise}\\
                       \end{array} \right . \text{,}\label{dens}
\end{equation}
where $A$ is an arbitrary constant. By introducing Eq. (\ref{dens}) into Eq. (\ref{Gamman2}), one obtains
\begin{equation}
 \Gamma(t)= \frac{A}{\omega_1-\omega_2}\left[\frac{\sin(\omega_1 t)} {t} -\frac{\sin(\omega_2 t)}{t} \right]\text{,} \label{Gammaf}
\end{equation}
with Laplace transform
\begin{equation}
\widetilde{\Gamma}(z)= \frac{A}{\omega_1-\omega_2} \left[\arctan\left(\frac{\omega_1}{z}\right)- \arctan\left(\frac{\omega_2}{z}\right)\right] \text{.}
\label{Gamzf}
\end{equation}

From the final-value theorem on Laplace transformation, note that
 \begin{eqnarray}
 \lim_{t\rightarrow\infty} R(t)&=& \lim_{z\rightarrow 0}z \widetilde{R}(z) = \lim_{z\rightarrow 0} \frac{z}{z+\widetilde{\Gamma}(z)}\notag \\
 &=& \frac{\omega_1 \omega_2}{A+\omega_1\omega_2}\neq 0, \label{Rtf}
 \end{eqnarray}
 which according with Khinchin theorem is non-ergodic~\cite{Lee07,Lapas08}.

\begin{figure}
\centering \includegraphics[width=0.5\linewidth]{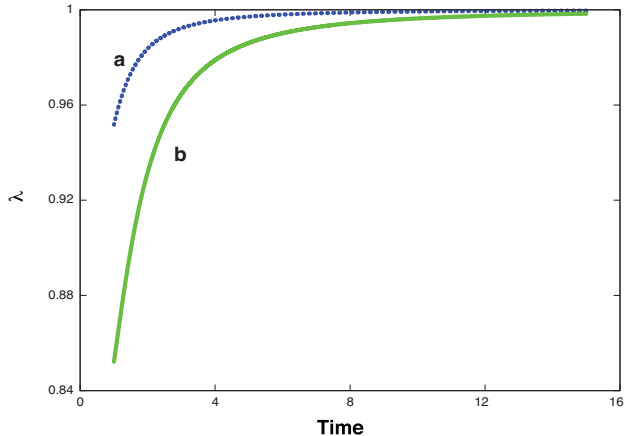}
\caption{\label{fig5} Scaling parameter $\lambda(t)$ as a function of time $t$.  We use the memory function, Eq. (\ref{Gammaf}), and obtain $\lambda(t)$ self-consistently as a map. ({\bf a}) $\omega_1=4$ and $\omega_2=3$. ({\bf b}) $\omega_1=2$ and $\omega_2=1$. For all curves we use $A=1$. Note that both curves converge rapidly for long times.}
\end{figure}

\begin{figure}
\centering \includegraphics[width=0.5\linewidth]{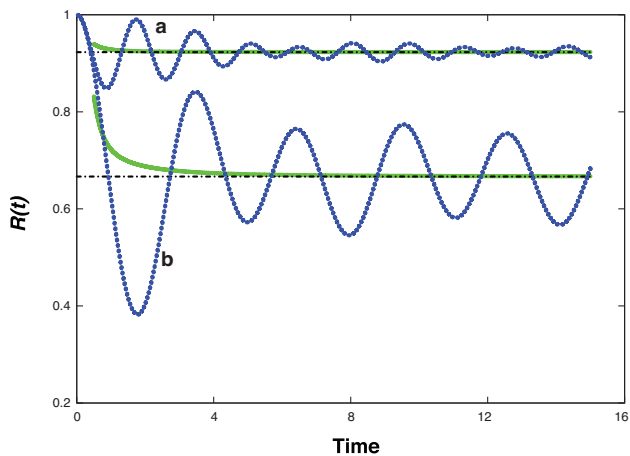}
\caption{\label{fig6} Correlation function $R(t)$ as a function of time $t$.  The numerical integration is performed over Eq. (\ref{eq6}) by using Eq. (\ref{Gamzf}). Data are the same as in Fig.~\ref{fig5}.}
\end{figure}

\begin{figure}[t!]
\centering \includegraphics[width=0.5\linewidth]{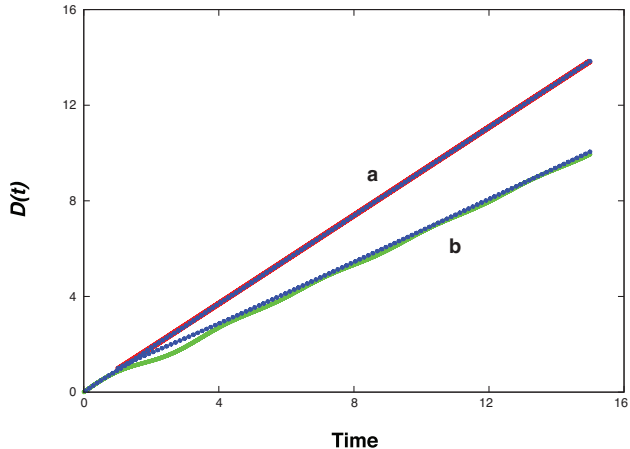}
\caption{\label{fig7} Diffusion coefficient $D(t)$ as a function of time $t$. Data are the same as in Fig.~\ref{fig5}. The oscillatory curves stand for the numerical results. The curves without oscillations are the analytical asymptotic limits. Note that the two curves collapse onto a single one in curve {\bf a}. }
\end{figure}

\begin{figure}
\centering \includegraphics[width=0.5\linewidth]{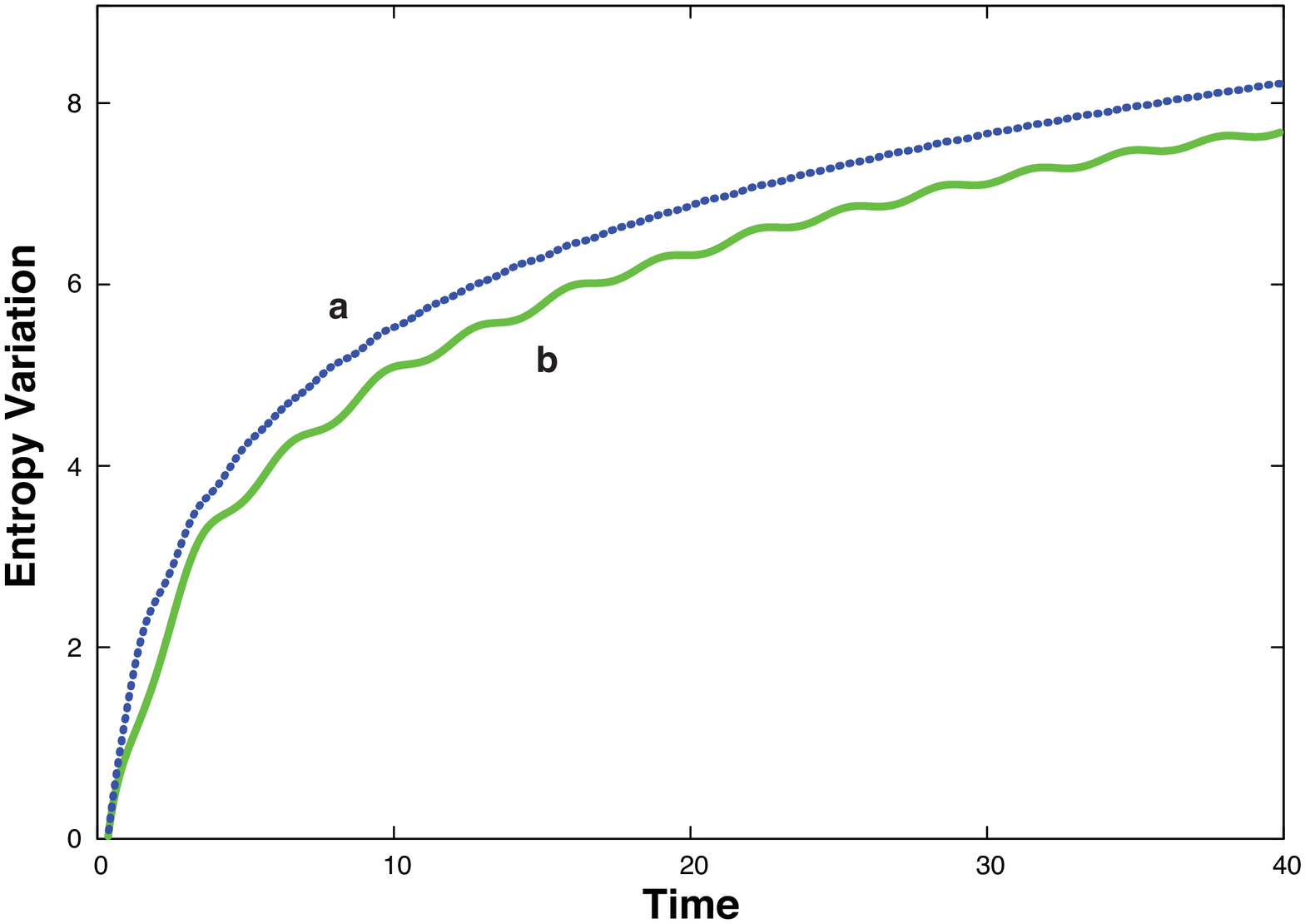}
\caption{(color online). Entropy variation $\Delta S(t)=S(t)-S(t_0)$, given by Eq. (\ref{SG}), as a function of time $t$, in arbitrary units. Here $t_0$ is the starting point, the data are as in Fig.~\ref{fig5}.}
\label{fig_entropy}
\end{figure}

Figure~\ref{fig5} displays the evolution of the parameter $\lambda(t)$ as a function of $t$. We obtain $\lambda$ computing iteratively $ \lambda_{n+1}(t)=\lambda(\lambda_n(t),\Gamma(z))$, using Eq. (\ref{zt}), and by introducing some arbitrary initial value  $\lambda_1$ (see details in Ref.~\cite{Ferreira12}). The iterative scheme continues until convergence is obtained; after 50 iterations $\vert \lambda_{n+1}(t)-\lambda_n(t)\vert < 10^{-14}$. As $t$ increases the function $\lambda(t)$ converges rapidly to one:  limit $\lim_{t\rightarrow \infty} \lambda(t) =1$. This corresponds to a ballistic (non-ergodic) diffusion.

In Fig.~\ref{fig6}, we show the behavior of  $R(t)$ as a function of time $t$. The wiggle line stands for the numerical integration results, the curves without oscillations are obtained by the scaling Eq. (\ref{zt}), and the horizontal line is the final value obtained from Eq. (\ref{Rtf}). One should note that the convergence of $R(t)$ is controlled by the convergence of $\lambda(t)$. From Figs.~\ref{fig5} and ~\ref{fig6}, one can see that while $R(t)$ is still oscillating, $\lambda(t)$ has already reached values very close to $1$. 

In Fig.~\ref{fig7}, we show the evolution of the diffusion coefficient $D(t)$ as a function of time. The fast convergence to the asymptotic limit obtained by the scaling is the reason for the small differences between the values of the diffusion coefficient observed in the figure. This shows again that $\psi(t)$ given in Eq. (\ref{Rho}) is a growing function of time.

In Fig.~\ref{fig_entropy} we plot the variation of Gibbs entropy  $\Delta S(t)=S(t)-S(t_0)$ from Eq. (\ref{eq:gibbs}) as a function of time, showing that entropy grows in any Gaussian process. The numerical data have been obtained as in Fig.~\ref{fig3}. We use $t_0=0.6$, and the data of previous figures. In (a) we have $\langle v^2(t_0) \rangle=1.426$ and $\langle x^2(t_0) \rangle=1.240$, and in (b) we have $\langle v^2(t_0) \rangle=1.401$ and $\langle x^2(t_0) \rangle=1.361$, the initial values are very close for both cases. The entropy increases due to diffusion is greater than the one related to the cooling process. In addition, we do not observe any oscillation in the entropy values such as we recognized in the temperature. The gas is subjected to oscillations in the temperature due to collective behavior. The violation of the second law never occurs, even for any time interval when heat flows from the cold reservoir to the hot gas; the absolute value of the entropy production in this process is smaller than the part produced in the diffusion one.

In our model, by considering the thermal reservoir of harmonic oscillators, one obtains the memory function in terms of the noise spectral density~\cite{Vainstein06a}, Eq. (\ref{Gamman2}). The internal complexity of the movement, which is explained here by the function $R(t)$, suggests the existence of a collection of relaxation variables. In fact, the relaxation time due to diffusive process may be different from one associated exclusively with the cooling process. In the case of the normal diffusion, either relaxation occurs fast and one recovers Newton’s cooling law. From a hypothetical memory function, resulting in the ballistic diffusion (non-ergodic), the effective relaxation time, Eq. (\ref{beta}), may exceed the experimental duration, similarly to ``fluctuation freezing" in silicate glasses~\cite{ Fabelinskii68,Gotze92, Rubin02}. Therefore, the relaxation parameters cannot reach their equilibrium values, remaining at an intermediate temperature between the initial temperature and the absolute temperature of the thermal reservoir~\cite{Lapas07}. The race between relaxation times makes the velocity correlation, and therefore the temperature, oscillate. In other words, the particles are subjected to oscillations in the temperature due to collective behavior, mainly due to the response mechanism of the thermal reservoir to the particle dynamics. This behavior can be explained in terms of the function $R(t)$ as well as of the relaxation time $\beta$, which can be a time-dependent parameter and exhibits a hierarchy of relaxation times~\cite{Madrid05}. Since R(t) converges to zero in thermodynamic equilibrium (or very close to equilibrium), all the diffusive regimes exhibit an entropy increase such as its rate of increase as well as its magnitude are limited to the values of $R(t)$. In the extreme case of ballistic diffusion, where $R (t)$ is nonzero for long time, entropy increases in a lesser rate than the other diffusive regimes. Anomalous behavior was reported in literature both in the velocity correlation function~\citep{Zwanzig70} as in time-dependent temperature measurements~\cite{Sampat94}.

\section{Conclusions} \label{sec:conclusions}
In this work, we have investigated the relaxation of the temperature of a system that undergoes anomalous diffusion and is in contact with a thermal bath. We have performed an analysis based on a generalized Langevin equation to derive an expression for the relaxation equation for the temperature of the system which resembles Newton's law of cooling but with a nonconstant coefficient that depends on the nature of the diffusion process. Temperature relaxation exhibits significant differences with respect to the case of normal diffusion in which Newton’s law of cooling holds. In particular, the temperature of the system may oscillate because of memory effects. 

Oscillations in Langevin dynamics in a somewhat different context were studied previously~\cite{Burov08,Jeon12,Kursawe13}. In another model, it has been shown that in systems with time-dependent and/or a spatially nonuniform temperature the diffusion is anomalous~\cite{Fulinski13}. In this context, anomalous diffusion can be detected in simple systems with nonstationary or/and nonuniform temperature; see also a recent review~\cite{Metzler14} for more details. A possible application of the temperature oscillation in systems exhibiting anomalous diffusion includes the coupling with Brownian motors, so that particles can be driven in a specified direction~\cite{Savelev03}.

In addition, we have shown that the entropy always increases, in such way that the diffusive process produces more entropy than the cooling process. The anomalous cooling law obtained could be used to analyze thermal effects in systems such as glasses, liquid crystals, biological cells, polymers, and so on,  in which diffusion is anomalous~\cite{Metzler99}. Also, the heat transport mechanism at the nanoscale, which allows on to go beyond the normal diffusion in solids~\cite{Ben-Abdallah13}. In addition, diffusion poses challenges in the understanding of fundamental concepts in statistical physics, such as the validation of the FDT~\cite{Costa03}, general properties of the correlation function~\cite{Vainstein06}, entropy~\cite{Lapas07}, ergodicity~\cite{Lapas08,Siegle10,Lee07,Burov10}, and Khinchin theorem~\cite{Lapas08,Lee07}. These memory effects could be more relevant in small-scale systems, since the reduction of the observational scales would lead to a modification of the interactions of the particles with the thermal bath and consequently to a time-dependent friction coefficient whose nature would depend on the immediate environment of the particles.

{\it Acknowledgments.}---This work was supported by the FAPDF, CAPES, and CNPq of the Brazilian Government and MICINN of the Spanish Government under Grant No. FIS2008-04386. J.M. Rubi acknowledges financial support from Generalitat de Catalunya under program ICREA Academia.

\end{document}